\documentclass[12]{article}

\newcommand{\ket}[1]{|#1\rangle}
\newcommand{\bra}[1]{\langle#1|}
\newcommand{\proj}[1]{\ket{#1}\bra{#1}}
\newcommand{\sprod}[2]{\langle#1|#2\rangle}

\begin{document}
\title{Impossibility of perfect quantum sealing of classical information.}
\author{H. Bechmann-Pasquinucci\thanks{Permanent address: {\rm 
UCCI.IT}, via Olmo 26, I-23888 Rovagnate, Italy}, G. 
M. D'Ariano,\\ 
C. Macchiavello\\
\small
Quantum Information Theory group (QUIT), Dipartimento di Fisica
``A.\,Volta'' and \\
\small
INFM - Unit\`a di Pavia,
Via Bassi 6, I--27100  Pavia, Italy
\footnote{url: {\tt {www.qubit.it}}}}
\date{January 14, 2005}
\maketitle
\begin{abstract}
Sealing information means making it publicly available, but with the possibility of knowing if it has been read. 
Commenting on [1],  we will show that perfect quantum sealing is not possible for perfectly retrievable 
information, 
due to the possibility of
performing a perfect measurement without disturbance, even on unknown states. 
The measurement is a collective one, and this makes the protocol of quantum sealing very interesting as the only 
example of the power of collective measurements in breaking security.

\end{abstract}
\section{Introduction}
Recently the idea of a quantum seal has been proposed\cite{Helle}. It was introduced as the quantum 
version of a classical seal. Classical seals is  what was used before the time of electronic transfer to 
seal important documents or letters, and it consists of a wafer of molten wax into which was pressed the 
distinctive seal of the sender. In the quantum version the seal becomes a way of encoding a classical message 
into quantum states. Ideally the 
quantum seal should possess exactly the same properties as the classical seal, however, as it will be shown 
here, some of the requirements are too restrictive to allow a perfect quantum mechanical solution. 

The paper is organized as follows. In Sections 2 and 3 we review the original ideas of quantum sealing. In Section 4 we present the impossibility proof for 
a certain class of quantum seals, including the original proposed scheme.
Finally, we summarize our results, and discuss alternative ideas 
in Section 5.

\section{The sealing protocol}
The classical wax seal has some very particular properties. First of all, it is important to realize that 
the sender of the sealed letter is not committed to the content, a new letter can always be written and 
sealed, and substitutes the previous one. The seal serves two purposes: it provides some kind of 
authentication to persons with prior knowledge of the symbol on the seal, and it indicates if the envelope 
has been opened. Notice furthermore that the seal can be broken by anyone who wishes to learn the content of 
the letter. The classical seal does not provide security for the content, but a way of knowing if the 
letter has been read.

The quantum seal as proposed in reference [1] works in the following way: Alice, who wants to write a classical
message and seal it with a quantum seal, encodes one bit of information into a product state of three qubits.
Two of the qubits, the message qubits, will be in the same state $\ket{0}$ or $\ket{1}$ (of the computational
(z)-basis), depending on whether Alice wants to encode bit value '0' or '1'. The third qubit, the control qubit,
will at random be prepared in one of the four states $\ket{0_x}$, $\ket{1_x}$, $\ket{0_y}$ or $\ket{1_y}$.
Notice that these six states together form the three mutually unbiased bases in two dimensions. In each triplet 
of qubits the position of the control qubit is for security chosen at random.

When Alice has written her full message, which is a product state of many qubits, she stores it in a quantum
memory in a publicly accessible place, and announces that the reading basis of the message is the z-basis. The
knowledge of the message reading basis allows everyone to read the message, because a measurement in the
$z$-direction followed by a simple majority vote on the results of each triplet will reveal the bit value 
encoded by Alice. For example, consider the following triplet of qubits $\ket{0}\ket{0_y}\ket{0}$, a 
measurement of each
single qubit in the $z$-basis will with equal probability give either '000' or '010', and a majority vote will
tell the reader that in either case the bit value encoded by Alice was '0'.  Notice that Alice can at any
moment check if the message has been read. This is due to the fact that a measurement in the $z$-basis will
change the state of the control qubits, and since Alice knows the position of each control qubit, she can check
if the state has been changed.

This far there is nothing which allows intended readers to verify if the message has been read. As in the 
classical case, this step requires that the reader has additional information. This problem is solved by 
allowing Alice to distribute copies of some of the qubits in the message among the intended readers, Bob-1,.. 
Bob-N. Each Bob will be given a small set of qubits in states corresponding to specific positions in the sealed 
message, and he will also be told the position of each of the qubits, but what is very important he will not be 
told the state! 

In order for Bob to check if the message has been read, he can borrow from the sealed message the qubits which
correspond to the copies Alice has provided him with, and on each pair he can perform a so-called
SWAP-test\cite{BCWW}. A SWAP-test allows to check if two states are identical without any knowledge of the
state, and moreover, if the states are identical they will not be disturbed by the test. This means that the
SWAP-test performed on an undisturbed sealed message will not inflict any errors.

The quantum sealing protocol as given above, contains practically all the desired features of a quantum seal: 
Alice is not committed, she can always write a new message and seal it. By revealing the reading basis, 
she enables the whole world to read her message, but by adding control qubits she can verify if someone has 
actually read it. Even more, by giving additional qubits to the intended readers  they too will be able to 
check if 
the message has been read, by performing the SWAP-test, without causing disturbance and without 
learning anything about 
the message.

\section{Reading without breaking the seal: collective measurements}

As was
already pointed out in the original paper on quantum seals\cite{Helle} the above protocol for 
quantum seals is only secure against single qubit attacks. Unfortunately it is not secure against collective 
attacks. Actually, by performing collective measurements
on each triplet of qubits (corresponding to one encoded classical bit), it is possible to learn the encoded
bit value without introducing any kind of errors. This is due to the fact that the twelve three qubit states
which encode the $0$ value are orthogonal to the twelve three qubit states which encode value $1$. Indeed
all the $0$-states lie in a subspace spanned by the following states (in the $z$-basis), $\ket{000}$,
$\ket{001}$, $\ket{010}$ and $\ket{100}$, whereas all the $1$-states lie in the orthogonal subspace spanned by
$\ket{111}$, $\ket{110}$, $\ket{101}$ and $\ket{011}$. This means that a measurement of the corresponding 
projectors:
\begin{eqnarray}
P^{(0)}&=&\proj{000}+\proj{001}+\proj{010}+\proj{100},\\
P^{(1)}&=&\proj{111}+\proj{110}+\proj{101}+\proj{011}
\end{eqnarray}
will distinguish perfectly  between the $0$ and the $1$ values --- without disturbing the state. This 
means 
that a collective measurement of this kind will allow the sealed message to be read without disturbance, 
which means to learn the message without introducing errors whence avoid detection.

Notice that the set of states used to encode the $0$ value is not orthogonal, and similarly for the value $1$, 
which 
means that it is impossible to distinguish among states within each set with certainty. Indeed,
someone performing collective measurements will be able to read the
sealed message without detection, because the measurement will reveal the {\it classical} bit value encoded in
the triplet of states, but the measurement will not reveal the quantum state. It is worth emphasizing again the 
collective nature of the measurement, that, as we'll see, is a general feature of our impossibility proof. 
Indeed, the sealing protocol provides a unique illustration of the power of collective measurements in breaking 
security.

\section{The impossibility proof}
Is it always possible to make perfect discrimination between two sets of states without disturbing
the measured system, even without a complete knowledge of the quantum state? This is essentially the
problem underlying the possibility of achieving a secure quantum protocol to seal classical
information. In fact, in order to make information publicly available, one needs to disclose publicly the
procedure of the quantum measurement which perfectly discriminates the encoded values of the logical
bit. On the other hand, the possibility of knowing that the seal has been opened needs a signature
left on the quantum system signaling that the system has been measured, namely the measurement
must produce a "disturbance" on the system.

The general scenario of the quantum protocol for sealing classical information is the following.
We know that the classical bit is encoded in two families of states, here denoted by
\begin{equation}
\ket{\psi^{(b)}_\lambda}\in{\cal H},~~ b=0,1,\label{famstates}
\end{equation}
where ${\cal H}$ is the global Hilbert space on which the logical bit is encoded, 
and $\lambda\in\Lambda$ is a running parameter labeling the states of the family.
The fact that the classical information is publicly available means that there is an openly known
POVM $\{P^{(b)}\}$ on ${\cal H}$ namely
\begin{equation}
P^{(b)}\ge 0,\; b=0,1,\qquad P^{(0)}+P^{(1)}=I_{{\cal H}}. 
\end{equation}
If we consider the situation of no reading-error for the classical information, then we must have
\begin{equation}
\bra{\psi^{(b)}_\lambda}P^{(b')}\ket{\psi^{(b)}_\lambda}
=\delta_{bb'},\qquad\forall\lambda\in\Lambda.
\label{POVM}
\end{equation}
Notice that, since $P^{(b')}$ is positive,  the square root of the operator is well defined and  Eq. 
(\ref{POVM}) 
is equivalent to
\begin{equation}
\bra{\psi^{(b)}_\lambda}\sqrt{P^{(b')}}\sqrt{P^{(b')}}
\ket{\psi^{(b)}_\lambda}=
\|\sqrt{P^{(b')}}\psi^{(b)}_\lambda\|^2=\delta_{bb'},\qquad\forall\lambda\in\Lambda,
\end{equation}
namely 
\begin{equation}
\sqrt{P^{(b')}}\ket{\psi^{(b)}_\lambda}=0,\;\hbox{for}\; b\neq b',\;\forall\lambda\in\Lambda,
\end{equation}
Upon defining the two Hilbert subspaces
\begin{equation}
{\cal H}^{(b)}={\rm Span}\{\ket{\psi^{(b)}_\lambda},\;\lambda\in\Lambda\},
\end{equation}
Eq. (\ref{POVM}) tells us that the POVM element $P^{(0)}$ must have support orthogonal to 
${\cal H}^{(1)}$, and $P^{(1)}$ must have support orthogonal to ${\cal H}^{(0)}$.
This also implies that ${\cal H}^{(0)}$ is orthogonal to ${\cal H}^{(1)}$, since, otherwise, there
would exist a common subspace whose elements, e. g. $\ket{\phi^{(0)}}=\ket{\varphi^{(1)}}$ would give  
\begin{equation}
\bra{\phi^{(0)}}P^{(1)}\ket{\phi^{(0)}}=\sum_{\lambda,\lambda'}
\sprod{\phi^{(0)}}{\psi^{(0)}_\lambda}
\bra{\psi^{(0)}_\lambda}P^{(1)}\ket{\psi^{(0)}_{\lambda'}}
\sprod{\psi^{(0)}_{\lambda'}}{\phi^{(0)}}=0,
\end{equation}
and analogously $\bra{\varphi^{(1)}}P^{(0)}\ket{\varphi^{(1)}}=0$. But, since 
$\ket{\phi^{(0)}}=\ket{\varphi^{(1)}}$ one also has that
$\bra{\varphi^{(1)}}P^{(1)}\ket{\varphi^{(1)}}=0$ (and
$\bra{\phi^{(0)}}P^{(0)}\ket{\phi^{(0)}}=0$), namely
$\bra{\varphi^{(1)}}P^{(0)}+P^{(1)}\ket{\varphi^{(1)}}=0$, which contradicts completeness.
In this way we have proved that ${\cal H}^{(0)}$ is orthogonal to ${\cal H}^{(1)}$, namely one has
the Hilbert space direct-sum decomposition
\begin{equation}
{\cal H}={\cal H}^{(0)}\oplus{\cal H}^{(1)}.\label{osum}
\end{equation}
On the other hand the direct-sum decomposition (\ref{osum}) implies that the two POVM elements 
must be orthogonal projectors, since they have orthogonal support and are complementary. 
If the two Hilbert subspaces ${\cal H}^{(0)}$ and ${\cal H}^{(1)}$ are not isomorphic, then we can
always extend the smallest one in such a way to make them so. Then, without loss of generality, we
can write the (extended) Hilbert space ${\cal H}$ as follows
\begin{equation}
{\cal H}={\cal H}^{(0)}\oplus{\cal H}^{(1)}\simeq{\cal H}^{(0)}\otimes {\bf C}^2.
\label{newtensor}
\end{equation}
In the logical-bit tensor product decomposition (\ref{newtensor}), the two families of states rewrite as follows
\begin{equation}
\ket{\psi^{(b)}_\lambda}\doteq\ket{\psi_\lambda}\otimes\ket{b}\in{\cal H},\qquad b=0,1,
\end{equation}
and the POVM is expressed as follows
\begin{equation}
P^{(b)}=I_{{\cal H}}\otimes\proj{b}.\label{LudPOVM}
\end{equation}
The POVM (\ref{LudPOVM}) can be achieved by the L\H{u}ders measurement
\begin{equation}
\rho\longrightarrow \frac{P^{(b)}\rho P^{(b)}}{{\rm Tr}[P^{(b)}\rho P^{(b)}]},  
\qquad p(b)={\rm Tr}[P^{(b)}\rho P^{(b)}],\label{Luders}
\end{equation}
where $p(b)$ denotes the probability of outcome $b=0,1$. Clearly the measurement (\ref{Luders})
allows to distinguish between the logical bits without disturbing the two families of states in Eq. 
(\ref{famstates}).

In the case of the protocol above we have
\begin{eqnarray}
{\cal H}^{(0)}&=&{\rm Span}\{\ket{000},\ket{001},\ket{010},\ket{100}\},\\
{\cal H}^{(1)}&=&{\rm Span}\{\ket{111},\ket{110},\ket{101},\ket{011}\},
\end{eqnarray}
and the embedding (\ref{newtensor}) can be achieved, for example by the unitary transformation
which just exchanges only the two states $\ket{011}$ and $\ket{100}$, with the logical qubit
played by the first physical one.

\section{Discussion and future developments}
We have shown that perfect quantum sealing of classical information is not possible if the information is 
perfectly retrievable,
namely the seal is insecure when classical information is encoded
into quantum states and classical error correction codes are used to ensure 
error free reading of the sealed
message. 
 This is due to the fact that the resulting quantum states representing the {\it logical} bit value (in
the above example three qubit states) become orthogonal, hence the quantum seal can be broken by a collective
measurement. We emphasize that the collective measurement can only
distinguish between the logical bit values, whereas the specific state remains unknown and undisturbed.
Quantum seals are a concrete example of the power of collective measurements in breaking security.

We want to point out that sealing classical information by means of classical error correcting codes into 
quantum states is not 
the only way of implementing to idea of quantum seals. Chau\cite{Chau} has presented a version of quantum 
seals which seals {\it quantum information} into quantum states by means of entanglement and quantum error 
correcting 
codes. Unfortunately the protocol works for 
one (or few) authorized verifier and not as was originally proposed, 
that all intended readers can verify that the seal is still intact. More recently, Singh and  Srikanth\cite{SS} 
have proposed to use quantum seals in combination with secret sharing, so that it is not the message which is 
sealed, but each share. In principle this allows for sealing both classical and quantum information. By sealing 
the shares and not the message they can avoid the problem which we have been addressing in this paper. 

Our impossibility proof forbids perfect security of sealing for perfect retrievability of classical information. 
However, it is in principle possible to restore security when the information cannot be recovered 
perfectly\cite{inpre}.

\section*{Acknowledgments}
This work has been supported by EC under project SECOQC (contract n. 
IST-2003-506813) and by Ministero della
Universit\`a e della Ricerca under Cofinanziamento 2003. 

\end{document}